\documentclass[12pt]{article}

\usepackage{graphicx}
\usepackage[centertags]{amsmath}
\usepackage{cite}
\usepackage{enumerate}
\usepackage{scrtime}
\usepackage{color}
\usepackage[sans]{dsfont} 
\usepackage{colortbl}
\usepackage{subfigure}
\usepackage{paralist}
\usepackage{lscape}
\usepackage{longtable}
\usepackage{ulem}
\usepackage{cancel}
\usepackage{dcolumn}
\usepackage{url}
\usepackage{bm}
\usepackage{amssymb}
\usepackage{axodraw}

\def\be{\begin{equation}}
\def\ee{\end{equation}}
\def\bea{\begin{eqnarray}}
\def\eea{\end{eqnarray}}

\makeatletter       

\renewcommand{\p@subsection}{}

\renewcommand{\p@subsubsection}{}

\makeatother

\def\ie{i.e., }

\newcommand{\alphagut}{\alpha_{\gut}}

\newcommand{\alphaprime}{\alpha'}

\newcommand{\gev}{{\rm ~GeV~}}
\newcommand{\gstring}[1]{g^{#1}_{\textsc{string}}}
\newcommand{\gut}{{\textsc{gut}}}
\newcommand{\lc}{\ell_{\textsc{c}}}
\newcommand{\lstring}{{\ell_{\textsc{s}}}}

\newcommand{\mgut}{{M_{\gut}}}
\newcommand{\mex}{{M_{\textsc{ex}}}}
\newcommand{\mc}{{M_{\textsc{c}}}}
\newcommand{\mpl}{M_{\textsc{pl}}}
\newcommand{\mstring}[1]{M^{#1}_{\textsc{string}}}

\def\tr{\mathrm{Tr~}}

\def\bctr{\begin{center}}
\def\ectr{\end{center}}

\def\mbthree{\mathbf{3}}
\def\mbthreeb{\mathbf{\overline{3}}}
\def\mbtwo{\mathbf{2}}

\newcommand{\SO}[1]{\ensuremath{\mathrm{SO}(#1)}}
\newcommand{\SU}[1]{\ensuremath{\mathrm{SU}(#1)}}
\newcommand{\U}[1]{\ensuremath{\mathrm{U}(#1)}}
\newcommand{\E}[1]{\ensuremath{\mathrm{E}_{#1}}}
\newcommand{\LieG}[1]{\ensuremath{\mathrm{G}_{#1}}}

\begin{document}

\begin{flushright}
OHSTPY-HEP-T-08-004\\
\end{flushright}

\begin{center}
{\huge On the string coupling in a class of stringy orbifold GUTs}
\vspace*{5mm} \vspace*{1cm}
\end{center}

\vspace*{5mm} \noindent \vskip 0.5cm \centerline{\bf Ben Dundee and
Stuart Raby} \vskip 1cm \centerline{ \it Department of Physics, The
Ohio State University,} \centerline{\it 191 W.~Woodruff Ave,
Columbus, OH 43210, USA} \vskip2cm

\begin{abstract}
In this short note, we examine the relationship between the string
coupling constant, $\gstring{}$, and the
grand unified gauge coupling constant, $\alphagut{}$, in a highly
successful class of models based on anisotropic
orbifold compactifications of the weakly coupled heterotic string.
These models represent a stringy embedding of SU(6) gauge-Higgs
unification in a five dimensional orbifold GUT. We find that the
requirement that the theory be perturbative provides a non-trivial
constraint on these models. Interestingly, there is a correlation
between the proton decay rate (due to dimension six operators) and
the string coupling constant in this class of models.
Finally, we make some comments concerning the extension of these
models to the six (and higher) dimensional case.
\end{abstract}

\newpage

String theory is potentially a theory of everything; however, it is
still an open question as to whether or not the standard model
is one of the effective field theories which lies in the
``landscape'' of possiblities.  The weakly coupled
$\E8 \otimes \E8$ heterotic string is an excellent framework for obtaining
effective low energy field theories with many of the
phenomenological properties of the minimal supersymmetric standard
model (MSSM). Most recently, several detailed ``benchmark" string
models with 3 families of quarks and leptons, and one pair of Higgs
multiplets have been obtained
\cite{Buchmuller:2005jr,Buchmuller:2006ik,Lebedev:2006kn,Lebedev:2006tr,Lebedev:2007hv}.
Additional vector-like exotics and $\U1$ gauge symmetries decouple;
the Yukawa couplings for quarks and leptons are
non-trivial and the top quark Yukawa coupling and GUT coupling are
equal at the GUT scale (i.e. $y_{top}(\mgut) = g_{\gut}$)
due to the property of gauge-Higgs
unification. In addition, the models have an exact R-parity, and a
$D_4$ family symmetry under which the two light families transform as a
doublet, and the Higgs and third family transform as singlets.
This latter property might be crucial for generating a hierarchy of
quark and lepton masses, while ameliorating the supersymmetric
flavor problem.

In a recent paper \cite{Dundee:2008ts} it was shown how this highly
successful class of string models
\cite{Buchmuller:2005jr,Buchmuller:2006ik,Lebedev:2006kn,Lebedev:2006tr,Lebedev:2007hv}
could accommodate gauge coupling unification in a 5D orbifold GUT
limit.  Given the exotic matter content of the two benchmark
models outlined in Reference \cite{Lebedev:2007hv}, we found (using
an effective field theory analysis) 252 ways to achieve unification
by varying the cutoff $\mstring{}$ in the effective field theory,
the compactification scale $\mc$, and (most importantly) the
spectrum of ``light'' exotics with mass $\mex$. Of the 252 different
solutions found, 48 were not already ruled out by current (dimension
six) proton decay bounds.  By assigning VEVs to MSSM singlets, we were
able to show how one could realize one of these solutions in the
``Model 1A'' of Reference \cite{Lebedev:2007hv}.  In addition, the
solution described in \cite{Dundee:2008ts} satisfies the constraints
for unbroken low energy supersymmetry: $F=D=0$.  This latter feature
is essential if we are to understand the origin of the hierarchy
between the electroweak and Planck scales.

In this paper we address the important question of whether any of
these constructions are consistent with a perturbative string
expansion.   We find a simple formula for the 10D string coupling
$\gstring{}$ (see Eqn. \ref{string_coupling_2}) and show that the
constraint $\gstring{} < 1$ is correlated with the longevity of the
proton.  Of course, this result applies only to a very small, even
minuscule, portion of the string landscape; however, the relevant
question is whether or not it is applicable to those very constrained
portions of the string landscape where the minimal supersymmetric
standard model can be shown to reside.

The models of Reference \cite{Lebedev:2007hv} are derived from an
orbifold compactification of the weakly coupled heterotic string:
formally $T^6/\mathbb{Z}_6$-II, which can be parameterized by the
root lattice $\LieG2 \times \SU3 \times \SO4$.  By varying the VEVs of
the $T$ (size) and $U$ (shape) moduli associated with the \SO{4}
lattice, it was shown in References \cite{Kobayashi:2004ud, Forste:2004ie,
Kobayashi:2004ya} that one can achieve a stringy embedding of the
highly successful orbifold GUT picture
\cite{Kawamura:2000ev,Hall:2001pg,Asaka:2001eh,Kim:2002im}.  In the
literature, this has been called ``anisotropic'' string
compactification \cite{Ibanez:1991zv,Ibanez:1992hc,Kobayashi:2004ud,
Kobayashi:2004ya,Hebecker:2004ce,Ross:2004mi}. The problem, of
course, is that the GUT coupling constant in the effective four
dimensional theory is proportional to the ten dimensional Yang-Mills
coupling (and thus the string coupling, $\gstring{}$) by a factor of
one over the volume of the six-dimensional compactification.  The
requirement of acceptable unification in the low energy effective
field theory may be
inconsistent with the requirement that the underlying string theory
be weakly coupled ($\gstring{} \lesssim 1$), depending on the
precise relationship between the two parameters.

By demanding that the underlying heterotic string theory still be
perturbative (\ie weakly coupled), we show how one can further
constrain the parameter space of our models---in fact, of the 252
solutions which were found in Reference \cite{Dundee:2008ts}, only
28 of them turn out to have $\gstring{} < 1$, see Table
\ref{tab:interesting_models} on page
\pageref{tab:interesting_models}.  Moreover, all of
these 28 models have a long lived proton, with $\tau(p\rightarrow
\pi^0 \ e^+) \gtrsim 10^{34}$ y.  Because the proton lifetime is
proportional to the fourth power of the compactification scale,
and the string coupling $\gstring{}$ is inversely proportional
to the volume of the compact space,
there is a correlation between $\mc$, $\gstring{}$ and $\mstring{}$,
which we make explicit.  This means that the question of weak string
coupling is not entirely decoupled from the low energy phenomenology
in these models.  In fact, for a reasonable choice of parameters, a
long lived proton seems to be \textit{synonymous with} weak string
coupling.  A particularly interesting detail is that the same
example which we constructed in Section 4 of Reference
\cite{Dundee:2008ts} will survive this round of scrutiny, with
$\gstring{}\sim 0.5$.  In addition, all but one of the nine models
which were categorized as ``interesting'' (see Table 9 in Reference
\cite{Dundee:2008ts}) are eliminated when we require the string
coupling to be small.  Thus the requirement that we be in a
perturbative regime of the underlying string theory gives a new,
non-trivial constraint on the ``mini-landscape'' models.  In light
of this requirement, we comment on the ability to interpret the
models of References
\cite{Buchmuller:2005jr,Buchmuller:2006ik,Lebedev:2006kn,Lebedev:2006tr,Lebedev:2007hv}
as six (and higher) dimensional orbifold GUTs.

\textit{The String Coupling.}---In a given string compactification,
the  string coupling is set by the VEV of a scalar field, called the
dilaton.  In general, one has $\gstring{2} \sim e^{2\phi}$.  In
order to find the exact relationship, one must start from the ten
dimensional effective action for the weakly coupled heterotic string
and compactify on some six dimensional manifold.  The four
dimensional effective action is \cite{Witten:1996mz}
\begin{equation}
   \mathcal{S}_{eff} = -\int d^{4}x \sqrt{g} e^{-2\phi} V_6
   \left\{\frac{4}{\alphaprime^4}R + \frac{1}{\alphaprime^3} \tr F^2 + \cdots\right\}.
\end{equation}
where $\phi$ is the (ten dimensional) dilaton, $V_6$ is the volume
of the compactification, and $\alphaprime$ is the parameter which
sets the string tension.  We can identify the coefficient of the
gravity term with Newton's constant:
\begin{equation} \label{newtons_constant}
   \frac{4 e^{-2\phi} V_6}{\alphaprime^4} \equiv
   \frac{1}{16 \pi G_N} \Rightarrow G_N \equiv \frac{\alphaprime^4 e^{2\phi}}{64\pi V_6},
\end{equation}
and the coefficient of the gauge kinetic term with the (four
dimensional) Yang-Mills coupling constant\footnote{Note that we have
normalized the gauge fields such that in the fundamental
representation of \SU{N} we have $\tr(T_a T_b) = \frac{1}{2}
\delta_{a b}$, which is the standard normalization used for
phenomenology.  In addition the GUT coupling $\alpha_{\gut}$ is
evaluated at the string scale $\mstring{}$.}:
\begin{equation} \label{gut_coupling}
   \frac{e^{-2\phi} V_6}{\alphaprime^3} \equiv \frac{1}{2g_{\gut}^2}
   \Rightarrow \alphagut \equiv \frac{\alphaprime^3 e^{2\phi}}{8\pi V_6}.
\end{equation}
The parameter $\alphaprime$ is related to the cutoff in the
effective field theory \cite{Ghilencea:2002ff,Dundee:2008ts}:
$\Lambda^{-2} \equiv\mstring{-2} \approx \alphaprime$. Note that
this parameter was chosen in such a way as to capture the maximum
amount of stringy (threshold) effects in the low energy effective
field theory without actually calculating them \cite{Ghilencea:2002ff}.
Of course, the exact relationship between $\alphaprime$ and $\mstring{}$
depends on the regularization scheme (see for example
\cite{Kaplunovsky:1987rp}). In particular, we will take the standard
definition of the \textit{string length} $\lstring$, such that it is
related to the cutoff by $\lstring \equiv
\frac{\sqrt{\alphaprime}}{2} \approx \frac{1}{2\mstring{}}$.
Finally, the compactification scale is given in terms of the radius
of the fifth dimension: $\lc = R \equiv \frac{1}{\mc}$.

By exploiting the duality between the $\E8 \otimes\E8$ heterotic
theory and heterotic-M theory, Hebecker and Trapletti argued
\cite{Hebecker:2004ce} that the proper relationship between the 10D
dilaton and the string coupling constant is given by\footnote{They
showed that for  $\gstring{} < 1$, the lowest lying massive state is
a perturbative heterotic string state, while for $\gstring{} > 1$ it
is a Kaluza-Klein mode of M theory.  At the present time, this is
the best estimate we know of for defining the perturbative heterotic
string regime.}
\begin{equation} \label{string_coupling}
   \gstring{2} \equiv \frac{8e^{2\phi}}{\left(2\pi\right)^7}.
\end{equation}
This gives us a relationship between the (four dimensional) GUT
coupling constant at the string scale and the string coupling. By
eliminating the dilaton dependence between Equations
(\ref{gut_coupling}) and (\ref{string_coupling}) we find
\begin{equation}
   \alphagut = \frac{\alphaprime^3 \left(2\pi\right)^6}{2^5 V_6}\gstring{2}
\end{equation}
Taking five directions compactified at the string length,
$\lstring$, and one direction compactified at $\lc$, we find
\begin{equation} \label{string_coupling_2}
   \gstring{2} = \alphagut \frac{\mstring{}}{\mc}.
\end{equation}
Note that it is entirely possible that the effective field theory be
weakly coupled, but that the underlying string theory be strongly
coupled.

\begin{figure}
        \includegraphics[height = 0.5\textheight]{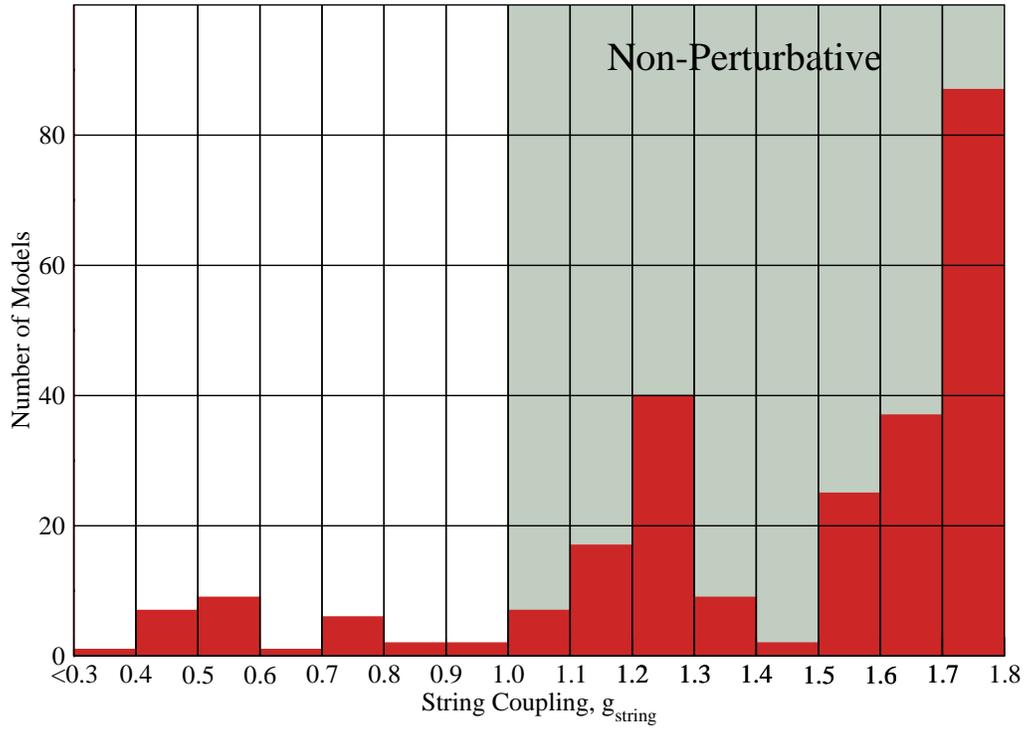}
        \caption{Histogram of the string coupling of the 252 solutions of Reference \cite{Dundee:2008ts}.
        Of the 48 models which were not eliminated previously because of dimension six proton decay,
        28 have $\gstring{} \lesssim 1$.}
        \label{fig:histogram}
\end{figure}

\textit{Are We Perturbative?}---Using the relationship in Equation
(\ref{string_coupling_2}), we can examine the 252 different
solutions found in Reference \cite{Dundee:2008ts}.  The results of
this analysis are shown in Figure \ref{fig:histogram}. Of the 48
models which were not eliminated previously because of dimension six
proton decay, 28 have $\gstring{} \lesssim 1$.  There is certainly a
preference for strong coupling in these models: this is a competing
effect between the ratio of the string scale to the Planck scale
(which sets $\alphagut{}$) and the ratio of the string scale to the
compactification scale (which sets $\gstring{}$).

In general, however, it is significant that only the
models with long lived protons have small string coupling.  As
discussed in  \cite{Dundee:2008ts} the proton lifetime scales as
$(\mc^4/\alphagut^2)$. Then using the relation between $\alphagut{}$
and the Planck scale
\begin{equation} \label{alphagut} \alphagut^{-1} = \frac{1}{8}
\left(\frac{\mpl}{\mstring{}}\right)^2
\end{equation} obtained by combining Equations
(\ref{newtons_constant}) and (\ref{gut_coupling}), and the dimension
6 operator contribution to the proton decay rate (see Reference
\cite{Dundee:2008ts}), we obtain the following useful formula for
the proton lifetime:
\begin{equation}
\tau(p\rightarrow \pi^0 e^+) \cong 5.21 \times 10^{40}
\left(\frac{\mc}{\mstring{}}\right)^4 \mathrm{yr}.
\end{equation}
We can then re-write Equation (\ref{string_coupling_2}) as
\begin{equation}
   \gstring{2} = \alphagut \left( \frac{5.21\times 10^{40}\mathrm{~yr} }{\tau(p\rightarrow
   \pi^0e^+)}\right)^{1/4}.
\end{equation}
The current (published) limit on the proton lifetime \cite{Yao:2006px}
\begin{equation} \tau(p\rightarrow e^+ + \pi^0) > 1.6
\times 10^{33} \rm{~yr}\end{equation}
implies that
\begin{equation}
   \gstring{2} \lesssim 600 \frac{\mstring{2}}{\mpl^2},
\end{equation}
where we have inserted the definition of $\alphagut$ in terms of the
Planck scale, Equation (\ref{alphagut}). If we take a typical value
for the string scale $\sim 5.0 \times 10^{17} \gev$ and the Planck
scale $\sim 1.2\times 10^{19} \gev$, we find that
\begin{equation}
   \gstring{2} \lesssim 1.
\end{equation}

Another interesting point is that the model described in Section 4
of Reference \cite{Dundee:2008ts} has a small string coupling.
There, we found $\mc \sim 2.2\times 10^{17} \gev$and $\mstring{}
\sim 1.0 \times 10^{18} \gev$.  We find $\gstring{} \sim 0.5$.
This is encouraging because we were able to show that that model is
consistent with $F=D=0$ and the decoupling of unwanted exotics from
the low energy spectrum.

Of the 48 solutions which we found in Reference
\cite{Dundee:2008ts}, we isolated a handful (9) which exhibited only
moderate hierarchies between the scales in the problem.  When we
look at the string coupling using Equation
(\ref{string_coupling_2}), however, we see that only one of them can
be derived from a model at weak coupling. Unsurprisingly, this is also the
model with the largest value of $\mc$ and thus the longest lived
proton.

Note that in all of the models with $\gstring{} < 1$, the
compactification scale $\mc$ is above or equal the 4D GUT scale:
see Table \ref{tab:interesting_models} on page
\pageref{tab:interesting_models}.
Hence the threshold corrections in these models, which focus the 3
low energy couplings, come predominantly from the contribution of
the exotics with mass $\mex$.  This result is particularly model
dependent.  While the KK modes contribute the universal power law
running which allows the theory to satisfy the weakly coupled
heterotic string boundary condition, Equation (\ref{alphagut}), they
also contribute to the differential running in a way which does not
focus the 3 gauge couplings.  It is the exotic matter at the
intermediate scale which furnishes a contribution to the
differential running, allowing for $\mc \gtrsim M_{\gut}$.  However,
it is possible that in other string models the KK modes alone would
be sufficient to both satisfy the weakly coupled heterotic string
boundary condition and focus the 3 gauge couplings.

Finally, we note that the models described in Reference \cite{Lebedev:2007hv}
can be interpreted as six dimensional orbifold GUTs.  If this is the
case, then the relationship in Equation (\ref{string_coupling_2})
will be amended:
\begin{equation}
   \gstring{2} = 2\frac{\mstring{2}}{M_5 M_6} \alphagut =
   16 \frac{\mstring{4}}{M_5M_6\mpl^2},
\end{equation}
where $\ell_{5(6)} \equiv M_{5(6)}^{-1}$ is the radius of the fifth
(sixth) direction.  In this case, it seems equally likely that a
weakly coupled model can be constructed.  If we take, for example,
$M_5 \sim M_6 \sim \mc$, and the typical value of $\mstring{} \sim
5\times 10^{17} \gev$, we find $\gstring{} \lesssim 1$ requires $\mc
\gtrsim 8 \times 10^{16} \gev$.

Taking more directions larger than the string length pushes us
toward stronger and stronger coupling, and it seems likely that if
this is the case then some other directions would have to be
\textit{smaller} than the string length.  This can be seen by
looking at the general relationship between $\gstring{}$ and the
other scales in the problem. If we take $n$ extra dimensions to be
large, we find
\begin{equation}
   \gstring{2} = 2^{n+2}\frac{\mstring{n+2}}{\mc^n\mpl^2}.
\end{equation}
If we take $n = 3$, and a typical string scale, we find that $\mc
\gtrsim 3\times10^{17} \gev$.

\textit{Conclusions.}---In this short note, we have analyzed the
string coupling in a class of highly successful models based on
anisotropic compactifications of the weakly coupled heterotic
string.  Of the 252 different solutions consistent
with gauge coupling unification found in Reference
\cite{Dundee:2008ts}, 48 were not already ruled out by current
(dimension six) proton decay bounds.  In this paper, out of the 48
solutions (not eliminated by the non-observation of proton decay) we
find 28 which are consistent with a weakly coupled heterotic string,
$\gstring{} < 1$ (see Figure \ref{fig:histogram}).

We also pointed out an interesting correlation between the string
scale, the Planck scale, and the compactification scale (which sets
the proton lifetime).  Specifically, a proton lifetime consistent
with current bounds on dimension six operators seems to require weak
coupling, for a reasonable choice of parameters.  Moreover, we were
able to show that one specific (and very well-motivated) example
\textit{does} require $\gstring{} \sim 0.5$.

For all cases with $\gstring{} < 1$, the compactification scale
$\mc$ is above or equal the 4D GUT scale, $\mgut \sim 3 \times
10^{16}$ GeV. Hence the threshold corrections in these models, which
focus the 3 low energy couplings, come predominantly from the
contribution of the exotics with mass $\mex$.  While the KK modes
contribute the universal power law running which allows the theory
to satisfy the weakly coupled heterotic string boundary condition,
Equation (\ref{alphagut}), they also contribute to the differential
running in a way which does not focus the 3 gauge couplings.  It is
the exotic matter at the intermediate scale which furnishes a
contribution to the differential running, allowing for $\mc \gtrsim
M_{\gut}$.  This result is model dependent and it is possible that
in other string models the KK modes alone would be sufficient to
both satisfy the weakly coupled heterotic string boundary condition
and focus the 3 gauge couplings.

Finally, we commented on extensions of this work to six (and higher)
dimensional orbifold GUTs---barring large threshold corrections from
somewhere else (\ie higher dimensional operators), it seems possible
to construct models which are consistent with the weak coupling
ansatz in six dimensions.  However, in going to higher dimensions,
it seems likely that one would have to look for models in which some
of the compact directions had radii \textit{smaller than} the string
length.

\textit{Acknowledgments.}---We would like to thank Ignatios Antoniadis
for a conversation that led to this work.  The authors are partially
supported under DOE grant number DOE/ER/01545-880.

\begin{landscape}
\begin{table}[ht!]
 \centering
 \caption{Subset of models listed in Reference \cite{Dundee:2008ts}
 which exhibit $\gstring{} \lesssim 1$.
 Note that we define $\vec{n}$ in terms of the brane localized exotics:
 $\vec{n}\equiv(n_3,n_2,n_1)$, where the following (brane-localized) matter
 gets mass at the intermediate scale $\mex$ :
 $n_3\times \left[(\mbthree,1)_{1/3}+(\overline{\mbthree},1)_{-1/3}\right]
 + n_2 \times \left[(1,\mbtwo)_{0}+(1,\mbtwo)_{0}\right] + n_1 \times
 \left[(1,1)_{1}+(1,1)_{-1}\right]$.  (See Reference \cite{Dundee:2008ts} for more details.)}
 \vspace{5mm}
 \label{tab:interesting_models}
 \begin{footnotesize}
 \begin{tabular}{c|c|c|c|c|c|c|c}
  \hline
  Bulk Exotics & $\vec{n}$&$\mstring{}$ in GeV & $\mc$ in GeV & $\mex$ in GeV &$\tau(p\rightarrow e^+ \pi^0)$ in yr & $\gstring{}$&$\alphagut^{-1}$\\
  \hline
  \hline
  None&$(4,2,0)$&$9.18\times 10^{17}$&$2.22\times 10^{17}$&$4.88\times 10^{13}$&$1.77\times 10^{38}$&$0.43$&22\\
      &$(2,1,0)$&$9.18\times 10^{17}$&$2.22\times 10^{17}$&$2.60\times 10^{9}$&$1.77\times 10^{38}$&$0.43$&22\\
      &$(3,2,3)$&$9.88\times 10^{17}$&$2.22\times 10^{17}$&$2.08\times 10^{9}$&$1.32\times 10^{38}$&$0.48$&19\\
      &$(4,3,6)$&$1.08\times 10^{18}$&$2.22\times 10^{17}$&$1.59\times 10^{9}$&$9.23\times 10^{37}$&$0.55$&16\\
      &$(4,2,1)$&$8.26\times 10^{17}$&$6.65\times 10^{16}$&$5.43\times 10^{13}$&$2.19\times 10^{36}$&$0.67$&27\\
      &$(4,2,2)$&$6.87\times 10^{17}$&$2.19\times 10^{16}$&$6.52\times 10^{13}$&$5.34\times 10^{34}$&$0.89$&39\\
      &$(2,1,1)$&$6.87\times 10^{17}$&$2.19\times 10^{16}$&$6.18\times 10^{9}$&$5.34\times 10^{34}$&$0.89$&39\\
      &$(3,2,4)$&$7.07\times 10^{17}$&$2.16\times 10^{16}$&$5.68\times 10^{9}$&$4.52\times 10^{34}$&$0.94$&37\\
      &$(4,3,7)$&$7.28\times 10^{17}$&$2.13\times 10^{16}$&$5.21\times 10^{9}$&$3.79\times 10^{34}$&$0.99$&35\\

  \hline
  $\left[(\mbthree,1)_{2/3,*} + (\mbthreeb,1)_{-2/3,*})\right]^{++}$ +  &$(4,3,1)$&$9.96\times 10^{17}$&$7.74\times 10^{17}$&$4.50\times 10^{13}$&$1.90\times 10^{40}$&$0.26$&19\\
  $\left[(1,\mbtwo)_{1,*} + (1,\mbtwo)_{-1,*})\right]^{--}$  & $(4,3,2)$&$9.73\times 10^{17}$&$2.22\times 10^{17}$&$4.61\times 10^{13}$&$1.40\times 10^{38}$&$0.47$&20\\
&$(2,2,2)$&$1.01\times 10^{18}$&$2.22\times 10^{17}$&$1.92\times 10^{9}$&$1.19\times 10^{38}$&$0.5$&18\\
&$(3,3,5)$&$1.12\times 10^{18}$&$2.22\times 10^{17}$&$1.43\times 10^{9}$&$7.97\times 10^{37}$&$0.58$&15\\
&$(4,4,8)$&$1.28\times 10^{18}$&$2.22\times 10^{17}$&$9.64\times 10^{8}$&$4.73\times 10^{37}$&$0.71$&11\\
&$(3,2,0)$&$8.79\times 10^{17}$&$6.55\times 10^{16}$&$5.10\times 10^{13}$&$1.61\times 10^{36}$&$0.75$&24\\
&$(4,3,3)$&$9.06\times 10^{17}$&$6.50\times 10^{16}$&$4.95\times 10^{13}$&$1.38\times 10^{36}$&$0.78$&23\\

  \hline
  $\left[(\mbthree,1)_{2/3,*} + (\mbthreeb,1)_{-2/3,*})\right]^{--}$ + &$(3,1,1)$&$1.01\times 10^{18}$&$2.22\times 10^{17}$&$1.92\times 10^{9}$&$1.19\times 10^{38}$&$0.5$&18\\
  $\left[(1,\mbtwo)_{1,*} + (1,\mbtwo)_{-1,*})\right]^{++}$&$(4,2,4)$&$1.12\times 10^{18}$&$2.22\times 10^{17}$&$1.43\times 10^{9}$&$7.97\times 10^{37}$&$0.58$&15\\

  \hline
  $\left[(\mbthree,1)_{2/3,*} + (\mbthreeb,1)_{-2/3,*})\right]^{++}$ +  &$(4,2,0)$&$9.73\times 10^{17}$&$2.22\times 10^{17}$&$4.61\times 10^{13}$&$1.40\times 10^{38}$&$0.47$&20\\
  $\left[(1,\mbtwo)_{1} + (1,\mbtwo)_{-1})\right]^{++}$&$(2,1,0)$&$1.01\times 10^{18}$&$2.22\times 10^{17}$&$1.92\times 10^{9}$&$1.19\times 10^{38}$&$0.5$&18\\
&$(3,2,3)$&$1.12\times 10^{18}$&$2.22\times 10^{17}$&$1.43\times 10^{9}$&$7.97\times 10^{37}$&$0.58$&15\\
&$(4,3,6)$&$1.28\times 10^{18}$&$2.22\times 10^{17}$&$9.64\times 10^{8}$&$4.73\times 10^{37}$&$0.71$&11\\
&$(4,2,1)$&$9.06\times 10^{17}$&$6.50\times 10^{16}$&$4.95\times 10^{13}$&$1.38\times 10^{36}$&$0.78$&23\\

  \hline
  $\left[(\mbthree,1)_{2/3,*} + (\mbthreeb,1)_{-2/3,*})\right]^{--}$ +  &$(2,1,0)$&$9.36\times 10^{17}$&$2.22\times 10^{17}$&$2.45\times 10^{9}$&$1.64\times 10^{38}$&$0.45$&21\\
  $\left[(1,\mbtwo)_{1} + (1,\mbtwo)_{-1})\right]^{--}$ &$(4,2,0)$&$9.36\times 10^{17}$&$2.22\times 10^{17}$&$4.79\times 10^{13}$&$1.64\times 10^{38}$&$0.45$&21\\
&$(3,2,3)$&$1.01\times 10^{18}$&$2.22\times 10^{17}$&$1.92\times 10^{9}$&$1.19\times 10^{38}$&$0.5$&18\\
&$(4,3,6)$&$1.12\times 10^{18}$&$2.22\times 10^{17}$&$1.43\times 10^{9}$&$7.97\times 10^{37}$&$0.58$&15\\
&$(4,2,1)$&$8.79\times 10^{17}$&$6.55\times 10^{16}$&$5.10\times 10^{13}$&$1.61\times 10^{36}$&$0.75$&24\\

  \hline
  \hline
 \end{tabular}
 \end{footnotesize}
\end{table}
\end{landscape}

\bibliography{ref}

\end{document}